\documentclass[aip,reprint,amsmath,amssymb,longbibliography]{revtex4-2}

\usepackage{amsmath,amssymb}
\usepackage{bm}
\usepackage{graphicx}
\usepackage[utf8]{inputenc}
\usepackage[T1]{fontenc}
\usepackage{mathptmx}
\usepackage{xcolor}
\usepackage{comment}
\usepackage{hyperref}

\hypersetup{
colorlinks=true,
urlcolor= blue,
citecolor=blue,
linkcolor= blue,
}

\definecolor{lightgray}{gray}{0.6}
\newif\ifptitle
\newif\ifpnumber
\newcounter{para}
\newcommand\ptitle[1]{\par\refstepcounter{para}
{\ifpnumber{\noindent\textcolor{lightgray}{\textbf{\thepara}}\indent}\fi}
{\ifptitle{\textbf{[{#1}]}}\fi}}

\definecolor{mblue}{RGB}{51, 77, 167}
\definecolor{mgreen}{RGB}{0, 131, 56}
\definecolor{mpurple}{RGB}{92, 47, 147}
\newcommand{\blue}[1]{\textcolor{mblue}{#1}}
\newcommand{\green}[1]{\textcolor{mgreen}{#1}}
\newcommand{\purple}[1]{\textcolor{mpurple}{#1}}

\newcommand{\appropto}{\mathrel{\vcenter{
  \offinterlineskip\halign{\hfil$##$\cr
    \propto\cr\noalign{\kern2pt}\sim\cr\noalign{\kern-2pt}}}}}

\newcommand{\mytitle}{Design and characterization of a low-vibration laboratory\\ with cylindrical inertia block geometry}

\begin{document}

\title{\mytitle}

\author{Wenjie Gong}
\affiliation{Department of Physics, Harvard University, Cambridge, Massachusetts 02138, USA}

\author{Yu Liu}
\affiliation{Department of Physics, Harvard University, Cambridge, Massachusetts 02138, USA}

\author{Wan-Ting Liao}
\affiliation{Department of Physics, Harvard University, Cambridge, Massachusetts 02138, USA}

\author{Joseph Gibbons}
\email[]{jgibbons@hga.com}
\affiliation{Wilson HGA Architects, 374 Congress St, Boston, Massachusetts, 02210, USA}

\author{Jennifer E. Hoffman}
\email[]{jhoffman@physics.harvard.edu}
\affiliation{Department of Physics, Harvard University, Cambridge, Massachusetts 02138, USA}
\affiliation{School of Engineering \& Applied Sciences, Harvard University, Cambridge, Massachusetts, 02138, USA}

\date{\today}

\begin{abstract}
Many modern nanofabrication and imaging techniques require an ultra-quiet environment to reach optimal resolution. Isolation from ambient vibrations is often achieved by placing the sensitive instrument atop a massive block that floats on air springs and is surrounded by acoustic barriers. Because typical building noise drops off above 120 Hz, it is advantageous to raise the flexural resonance frequencies of the inertia block and instrument far above 120 Hz. However, it can be challenging to obtain a high fundamental frequency of the floating block using a simple rectangular design. Here we design, construct, and characterize a vibration isolation system with a cylindrical inertia block, whose lowest resonance frequency of 249 Hz shows good agreement between finite element analysis simulation and directly measured modes. Our simulations show that a cylindrical design can achieve higher fundamental resonance frequency than a rectangular design of the same mass.
\end{abstract}

\maketitle

\section{\label{sec:Intro}Introduction}

\ptitle{STM capability \& noise requirements}
Revolutionary advances in atomic-scale fabrication and measurement equipment have required parallel advances in vibration mitigation technology. Among atomic-scale tools, the scanning tunneling microscope (STM) remains one of the most versatile and popular, but also the most sensitive to vibrations. An STM can be used to image the topographic shape of a surface with sub-atomic precision, to measure both filled and empty electronic density of states, to quantify momentum-resolved electronic band structure via quasiparticle interference imaging,\cite{GeRSI2019} and to manipulate individual atoms into custom configurations.\cite{CelottaRSI2014} These operations all require precise positioning of an atomically sharp tip within a nanometer of the sample of interest. Furthermore, STM measurements rely on quantum mechanical tunneling of electrons between tip and sample -- a process that is exponentially sensitive to the tip-sample separation $z$.\cite{BinnigAPL1982} Specifically, a 1 \AA\ change in $z$ typically corresponds to an order of magnitude change in tunneling current. Cutting-edge experiments call for current noise $\delta I < 1$ pA and position noise $\delta z < 1$ pm.\cite{GeRSI2019}

\begin{figure}[b]
\includegraphics[width=\columnwidth]{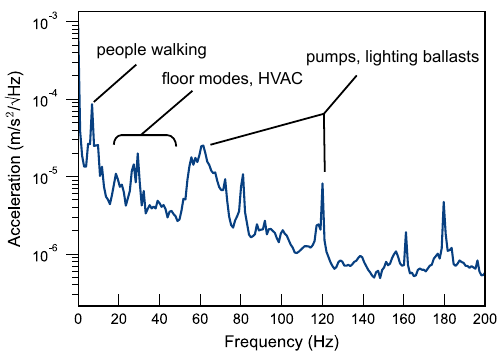}
\caption{Vertical acceleration spectrum, measured on the basement foundation of the Laboratory for Integrated Science \& Engineering at Harvard University, the floor on top of which our inertia block is situated. Typical building vibration noise peaks below 120 Hz, with few significant contributions at higher frequencies.}
\label{fig1}
\end{figure}

\ptitle{Typical building noise}
The unmitigated vibration spectrum measured on the basement floor of our laboratory is shown in Fig.\ \ref{fig1}, and is reflective of typical sources of building noise. Human footsteps excite the floor around 1-3 Hz, while the frame, walls, and floor of a building have shear and bending modes typically between 15 and 25 Hz.\cite{PohlIEEE1986} Heating, ventilation, and air conditioning (HVAC) systems contribute a rumbling noise from 10 to 30 Hz, while nearby traffic can add noise from 20 to 60 Hz.\cite{Leventhall1988} Many building systems such as pumps and fluorescent lighting ballasts effectively rectify the 60 Hz power, thus vibrating at 120 Hz. To protect against these noise sources, costly ultra-low-vibration facilities have been constructed for nanofabrication and imaging research around the world.

\begin{figure*}
\includegraphics[width=\textwidth]{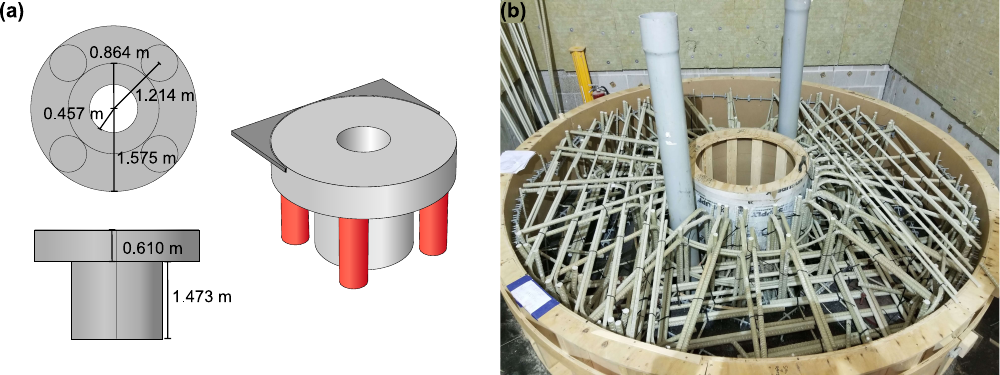}
\caption{(a) Dimensions of our cylindrical inertia block. (b) Photograph of the fiberglass rebar to be cast inside the concrete cylindrical block.} \label{fig2}
\end{figure*}

\ptitle{Low-vibration lab design}
The standard strategy for low-vibration facility design is to realize an effective low-pass filter for the building's ambient vibrations, with fastest possible rolloff, to avoid exciting the flexural modes of the sensitive instrument of interest, which are typically engineered to high frequency. For example, the relative tip-sample displacement modes of the STM are typically in the kHz range.\cite{PohlIEEE1986, OkanoJVSTA1987, Chen2007, AstRSI2008} Some facilities employ a separate, smaller base slab within the building, cut off from the main foundation, to reduce the transmission of building noise;\cite{IwayaRSI2011, MachidaRSI2018, BattistiRSI2018} however, coupling through a narrow construction joint may be hard to predict. A more controlled solution employs one or more layers of massive inertia block(s), supported by pneumatic isolators or active dampers that act as soft springs to achieve a low rolloff frequency for vertical movement.\cite{AmickNCEJ1998, WeaverJIEST2009, SongRSI2010, LortscherNanoscale2013, AssigRSI2013, MisraRSI2013, MacLeodRSI2016, KimRSI2017, VoigtlanderRSI2017, VonAllwordenRSI2018} However, either separate base slabs or floating blocks may have flexural modes of their own that lie within the frequency range of typical building excitations.\cite{AmickNCEJ1998, LortscherNanoscale2013, MacLeodThesis2013} Several researchers have noted that the relatively small mass of these blocks, compared to the larger building foundation, can increase their susceptibility to building vibrations, thus inadvertently magnifying noise transmission to the STM in some cases.\cite{AlbersRSI2008, IwayaJVSTA2012}

\ptitle{Rectangles have low $f_0$, here we do cylinder}
For effective vibration noise isolation, the inertia block must have fundamental flexural mode frequency $f_1$ far above the highest significant building excitation, i.e.\ $f_1 \gg 120$ Hz. However, such a high resonance frequency is difficult to achieve with a typical massive block of rectangular shape, particularly when the block thickness is limited by the space constraints of an existing building. Here we develop the first inertia block with a cylindrical shape, and demonstrate its lowest flexural mode above 240 Hz. We find reasonable agreement between finite element analysis simulations and experimental measurements of both the frequencies and shapes of the lowest four excited modes. Our work shows that for a space-constrained block with fixed mass, the cylindrical shape is the optimal choice to maximize flexural mode frequencies.

\section{\label{sec:Build}Block construction}

\ptitle{Cylindrical block dimensions}
The dimensions of our new cylindrical block design are shown in Fig.\ \ref{fig2}(a). The diameter was constrained to 3.15 m, and the total height was constrained to $\sim2$ m by the dimensions of the pre-existing room. The block is supported by four symmetrically-placed pneumatic isolators (model PD1001H from Integrated Dynamics Engineering), each with nominal vertical resonance frequency of $f_0 = 0.97$ Hz. To achieve this low $f_0$, the pneumatic isolator height further constrains the cylinder rim height to $\sim0.6$ m.

\ptitle{Cylindrical block construction}
To manufacture the block, cylindrical formwork was custom fabricated to fit within the lab, and installed over a bond breaker that allowed the block to be cast directly on the slab. We used fiberglass instead of more traditional steel rebar, to avoid coupling with a large magnet that will be part of our future experiment. The fiberglass reinforcement rods (Aslan 100 from Owens Corning, with diameters ranging from 0.5 to 1.0 inches) were assembled as bundles of individual loops, placed within the formwork at four inch intervals, and held in place with plastic ties, as shown in Fig.\ \ref{fig2}(b). Concrete was placed through a single pour to avoid structural cold joints.

\ptitle{Composite block properties}
Several additional samples of pure concrete from the same pour were placed beside the block. After 28 days of curing, three of these samples were tested and found to have mass density $w=144$ lbs/ft$^3$ and average compressive failure strength $f_c'=5077$ psi, which surpassed the design specification of 4000 psi. Using Pauw's empirical formula,\cite{Pauw1960} we compute Young's modulus of the concrete as $Y=33 w^{3/2} \sqrt{f_c'} = 4030\;\mathrm{ksi}=28.0$ GPa. Meanwhile, the fiberlass has mass density 134 lbs/ft$^3$ and Young's modulus 46 GPa, and makes up 4.0\% of the block by volume. From the weighted average of fiberglass and concrete, we estimate the effective composite properties of the block, mass density 2300 kg/m$^3$ and Young's modulus 28.7 GPa.

\begin{figure*}
\includegraphics{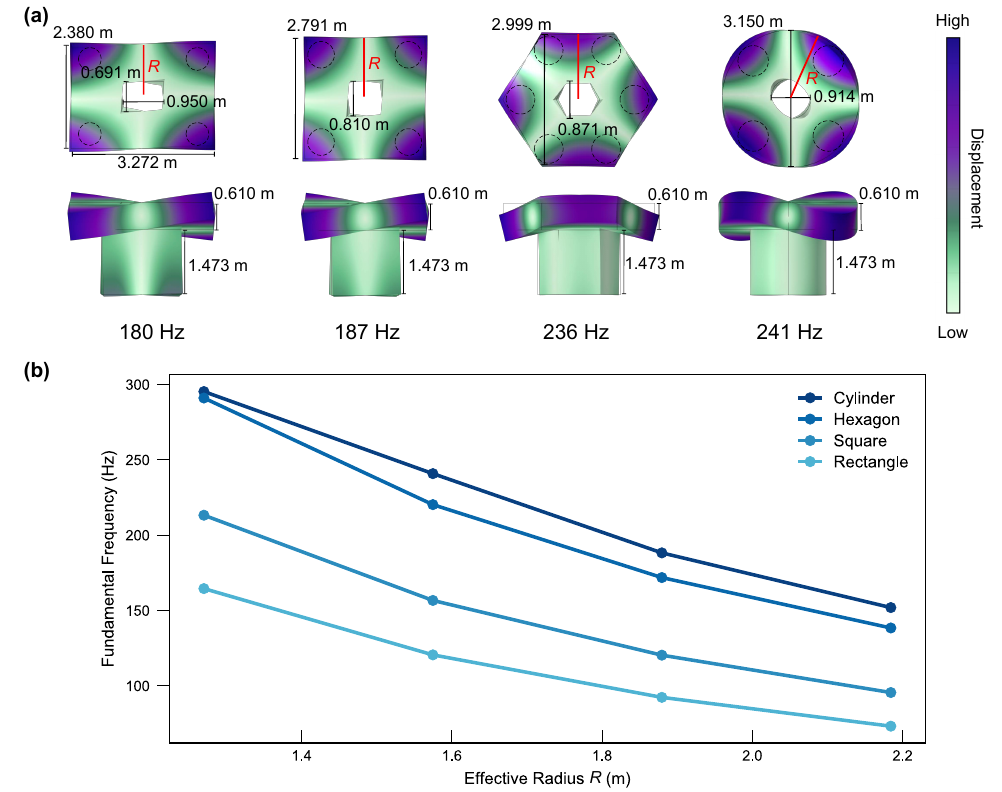}
\caption{(a) Modeled fundamental flexural modes of a set of four inertia blocks with the same mass but varying shape. All blocks have the same height and were scaled laterally as needed to maintain a volume of 6.83 m$^3$ (corresponding to mass 15,700 kg). All blocks were modeled with mass density 2300 kg/m$^3$, Young’s modulus $2.87\times10^{10}$ Pa, and Poisson’s ratio 0.20. The cylindrical model on the right matches the dimensions of our newly constructed block, shown in Fig.\ \ref{fig2}. (b) Frequency of the lowest flexural mode, for a set of four concrete block shapes, as a function of block size $R$, as defined in red on panel (a).}
\label{fig3}
\end{figure*}

\section{\label{sec:Sim}Simulating resonance modes}

\ptitle{General model parameters}
We simulated the flexural modes of a variety of inertia blocks using the commercial finite element analysis package {\sc comsol multiphysics 5.4}. In practice, the allowed modes are partially constrained by the contacts with the pneumatic isolators. We simulated this constraint by creating circular domains representative of each isolator's contact area on the underside of each block. These circular surfaces were set to spring boundary conditions, with total effective spring constant $k=(2 \pi f_0)^2M$, where $f_0=1$ Hz is the resonance frequency of the isolator and $M$ the mass of the block. The remainder of each block surface was set to free boundary conditions. We used the composite mass density 2300 kg/m$^3$ and Young's modulus $2.87\times10^{10}$ Pa computed in Section \ref{sec:Build}, with Poisson's ratio 0.20.

\ptitle{Controlled simulation of block shape}
A controlled test of inertia block shape is shown in Fig.\ \ref{fig3}(a), where we modeled the fundamental frequency of our cylindrical block along with blocks of identical mass but different shape. When mass, composition, and number of isolators are standardized but shape is varied, blocks with higher symmetry result in higher fundamental frequencies, with the cylindrical block having the highest predicted fundamental frequency of 241 Hz. In Fig.\ \ref{fig3}(b), we show how the fundamental frequency of decreases with block size, for each shape.

\ptitle{Simulate higher modes of cylinder}
The first four relevant simulated flexural modes of the cylindrical inertia block are shown in Fig.\ \ref{fig6}(a). Here we have omitted several modes with insignificant vertical displacement of the block's upper face, as these modes would not substantially affect the motion of the STM, but a complete set of simulated modes up to 800 Hz is shown in the supplementary material, Fig.\ \ref{fig:all-sim-modes}. The simulations provide only an estimate of the frequencies and mode shapes, because the modeled composition of the inertia block was oversimplified. In reality, fiberglass bars form a structured frame around which the concrete of the block is set. But in the simulation, only homogeneous weighted-average properties of concrete and fiberglass were used. To test the accuracy of this simple model, we must measure the block modes experimentally.

\section{\label{sec:Mes}Measuring resonance modes}

\begin{figure}
\includegraphics{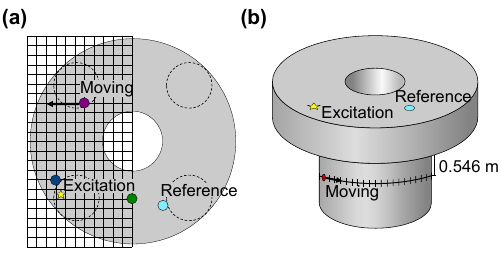}
\caption{(a) A diagram of the experimental setup for measuring vertical displacement response on the face of the block. (b) A diagram of the experimental setup for measuring horizontal displacement response on the side of the tail of the block. Excitation and reference points are consistent between (a) and (b).}
\label{fig4}
\end{figure}

\ptitle{Measurement locations}
Experimentally determining the relevant resonance modes of the cylindrical inertia block requires measuring the vertical displacement response of the top of the block at each excitation frequency. To delineate the measurement locations, we overlaid a grid with spacing 15 cm on half of the top face, as shown in Fig.\ \ref{fig4}(a). We thus measured a total of 165 data points for half the block, and used 180$^{\circ}$ rotational symmetry to populate the data for the other half of the block. For a more complete understanding of the mode shapes, we also measured the lateral displacement of the tail of the block, at the 40 points shown in Fig.\ \ref{fig4}(b), corresponding to $9^{\circ}$ angular resolution. Here, unlike the top face of the block, we directly measured all 40 locations on the circumference, and did not use symmetry to fill in the data.

\begin{figure}
\includegraphics[width=\columnwidth]{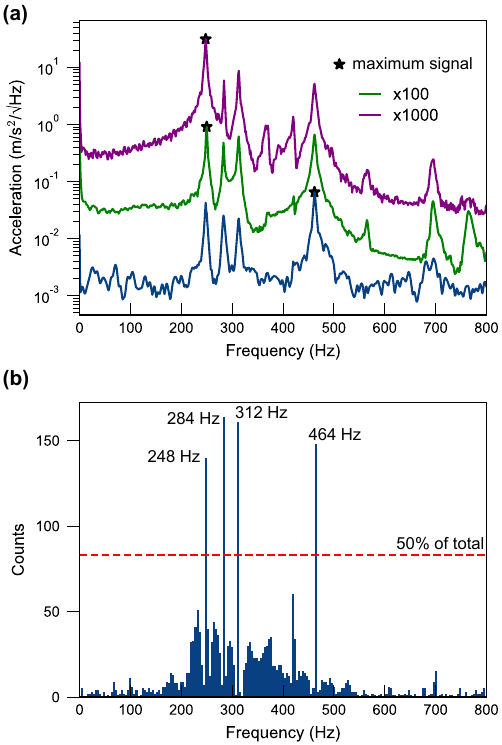}
\caption{(a) Typical acceleration spectra measured at three locations on the top face of the block. Each spectrum shown is the average of 10 FTs of consecutive 1-second data intervals, before normalization by reference. The green and purple spectra are vertically offset for clarity, by factors of 100 and 1000, respectively. (b) The distribution of frequencies at which peaks occurred in the acceleration spectra taken across all points on the face of the block. More than half of all acceleration spectra had peaks within 2 Hz of 248 Hz, 284 Hz, 312 Hz, and 464 Hz. }
\label{fig5}
\end{figure}

\ptitle{Top measurement procedure}
To measure the resonance frequencies, we first excited the block by impact with a rubber mallet while standing on the stationary floor adjacent to the block. The mallet effectively drove the block at all frequencies up to $\sim500$ Hz. Within tens of ms of the excitation, the vertical acceleration spectrum was measured at a given grid point on the top face of the block with a Wilcoxon 731a accelerometer. A reference accelerometer, also a Wilcoxon 731a, was kept at a constant location outside the grid of data points to normalize the amplitude of each mallet strike. The signals from both accelerometers were read simultaneously by a Stanford Research 785 spectrum analyzer for a 1 s time span after each excitation, and the Fourier transform (FT) of both signals was computed with a resolution of 1 Hz and range of 800 Hz. For each grid point, both accelerometer responses were averaged over 10 excitations. Typical spectra from 3 different positions are shown in Fig.\ \ref{fig5}(a).

\begin{figure}[!ht]
\includegraphics{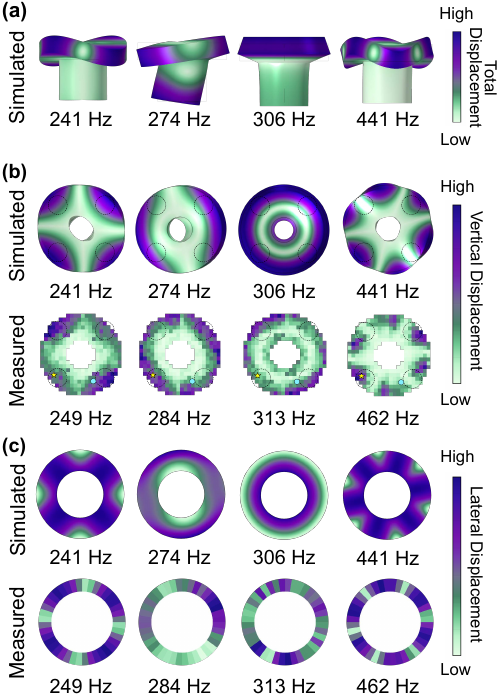}
\caption{(a) The first four simulated flexural resonance modes that give rise to significant $z$ displacement of the top surface of the cylindrical block. The purple-green colorscale shows total displacement, $\sqrt{(dx)^2+(dy)^2+(dz)^2}$. (b) Simulated top view of each mode, displayed on the first row, with the corresponding measured mode on the second row. Here, the purple-green colorscale shows vertical displacement, $dz$. (c) Simulated cross-section of the tail of the block at each mode, displayed on the first row, with the corresponding measured mode on the second row. Here, the purple-green colorscale shows horizontal displacement, $\sqrt{(dx)^2+(dy)^2}$.}
\label{fig6}
\end{figure}

\ptitle{Tail measurement procedure}
We used the same excitation and collection procedure to measure the lateral motion of the block's tail. A Wilcoxon 731a accelerometer was mounted sideways with double-stick tape on the tail of the block to measure the horizontal acceleration at each point on the circumference. During all vertical and horizontal measurements shown in Figs.\ \ref{fig5} and \ref{fig6}, both excitation point and reference point were held constant, and the inertia block was floated, with the pneumatic isolators pressurized to 80 psi. The door of the facility was left open and no attempt at acoustic isolation was made. Two analogous sets of measurements, with the block unfloated, and with a second excitation point, are shown in the supplementary material, Fig.\ \ref{fig:second-data}.

\ptitle{Spectrum data analysis}
We extracted the relevant flexural mode frequencies from the measurements on the top face of the inertia block, because this vertical displacement has most impact on the crucial tip-sample junction of the STM, and because the 165 measurements on top constituted a larger dataset than the 40 lateral points. We set a threshold acceleration response by finding the maximum signal from each of the 165 raw FTs (excluding the 0 Hz point), then taking 5\% of the median of these maxima. The frequencies at which peaks greater than this threshold occurred were extracted from each FT, and plotted in a histogram ranging from 0-800 Hz with bin width 4 Hz in Fig.\ \ref{fig5}(b). Only four bins had an occurrence count greater than 83 out of 165, or half of the total number of points. The midpoint frequencies of each of these four bins were taken to approximate the first four resonance frequencies of the inertia block. A Lorentzian fit was then conducted to find peaks within $\pm10$ Hz of each of these four frequencies across all grid points, and the centers of each fit peak were averaged to yield the more precisely measured resonance frequencies of 249 Hz, 284 Hz, 313 Hz, and 462 Hz. These measured frequencies track closely the simulated frequencies of 241 Hz, 274 Hz, 306 Hz, and 462 Hz shown in Fig.\ \ref{fig6}, despite our crude approximation of a homogeneous, isotropic cylinder with weighted-average properties of concrete and fiberglass. We note that the measurements are consistently $\sim2-4\%$ higher than simulations, which is to our advantage.

\ptitle{Quality factor} The Lorentzian width around each of the first four measured flexural mode frequencies yields an average quality factor of $Q_{\mathrm{concrete}} =112\pm34$, which falls at the stiff end of the typical range for normal concrete.\cite{AmickSPIE2005damping,Bachmann1995} (For fit details, see the supplementary material, Fig.\ \ref{fig:lorentzianfits}.) This relatively high quality factor, which determines the susceptibility of the block to amplify environmental noise at the mode frequencies, emphasizes the importance of increasing the fundamental flexural mode far above the building noise sources. The possibility to reduce $Q_{\mathrm{concrete}}$ by modifying the concrete with polymers has also been discussed.\cite{AmickSPIE2005damping}

\ptitle{Mode shape data analysis}
We compare the simulated and measured flexural mode shapes in Fig.\ \ref{fig6}. For each of the four measured peak frequencies $f_i$, we plotted the spatial shape of the corresponding mode by extracting the amplitude of the acceleration spectrum at that frequency from each grid point on the block, $a_{\mathrm{grid}}(f_i)$. We normalized each amplitude by dividing by the corresponding amplitude of the simultaneously-acquired reference data at the same frequency, $a_{\mathrm{ref}}(f_i)$, to account for variations in the excitation.  The normalized vertical measurements on the block's top face are shown in the second row of Fig.\ \ref{fig6}(b), and the normalized lateral measurements on the block's tail at the same four frequencies are shown in the second row of Fig.\ \ref{fig6}(c). The measured and simulated shapes of these four modes are in good agreement. Notably, all of our measured modes have nodes near the center of the block and antinodes near the edges. As the STM will rest on a table placed near the center of the block, displacement of the edges of the block is less of a concern.

\section{\label{sec:Mes}Conclusions}

\ptitle{Higher symmetry is better}
In designing an ultra-low-vibration laboratory facility, we aimed to maximize the fundamental flexural mode frequency of a massive inertia block, in order to minimize its susceptibility to typical dominant building noise sources in the 0-120 Hz range. We have thus developed the first cylindrical reinforced-concrete inertia block for use in vibration-sensitive research. Our simulations of various block shapes with controlled mass and composition show that higher symmetry leads to higher frequency flexural modes.

\ptitle{New lateral measurement}
We presented a systematic method for measuring both the vertical \textit{and} lateral components of the flexural modes of an inertia block. Previous measurements focused on the vertical component of block motion,\cite{MacLeodThesis2013} because it couples most directly to the typical orientation of the exponentially-sensitive tip-sample junction of an STM. But understanding the lateral components of the flexural modes remains important, because the three-dimensional block modes can couple lateral noise -- either building vibrations or acoustic excitations -- into vertical displacement of the top surface. For instance, an analysis of the piezo scanning element in an STM has shown that a driving signal can generate super- and sub-harmonic frequency peaks that couple horizontal and vertical motion of the piezo.\cite{RostAJC2009}

\ptitle{Simple approximations work}
Comparison between simulated and experimental resonances shows self-consistency in both mode shapes and frequencies, thus justifying both the model and the measurement technique. The complex internal structure of a reinforced-concrete inertia block can be reasonably approximated by assuming a uniform composition with weighted-average mass density and Young's modulus, such that simple finite element analysis simulations give good estimates within $\sim2-4\%$ of the measured resonance frequencies. With this demonstrated ease and accuracy of modeling, computer simulations can be effectively employed in the design process of an inertia block to ensure that the block’s fundamental frequency lies well above building noise sources.

\begin{acknowledgments}
We acknowledge funding support from the Harvard College Research Program. We thank Jeffrey Zapfe of Acentech, David Tuckey of LeMessurier, and Ben MacLeod for helpful conversations.
\end{acknowledgments}


\section*{\label{sec:Refs}References}

\providecommand{\noopsort}[1]{}\providecommand{\singleletter}[1]{#1}%

\clearpage
\setcounter{table}{0}
\renewcommand{\thetable}{S\Roman{table}}
\setcounter{equation}{0}
\renewcommand{\theequation}{S\arabic{equation}}
\setcounter{figure}{0}
\renewcommand{\thefigure}{S\arabic{figure}}
\setcounter{section}{0}
\renewcommand{\thesection}{S\arabic{section}}

\onecolumngrid
\noindent \textsf{\textbf{Supplemental Material for:}}

\vspace{3mm}

\noindent {\Large \textsf{\textbf{\mytitle}}}

\vspace{3mm}

\textsf{Wenjie Gong, Yu Liu, Wan-Ting Liao, Joseph Gibbons, Jennifer E.\ Hoffman}

\section{Simulated Modes}

\begin{figure}[h!]
    \centering
    \includegraphics[width=\columnwidth]{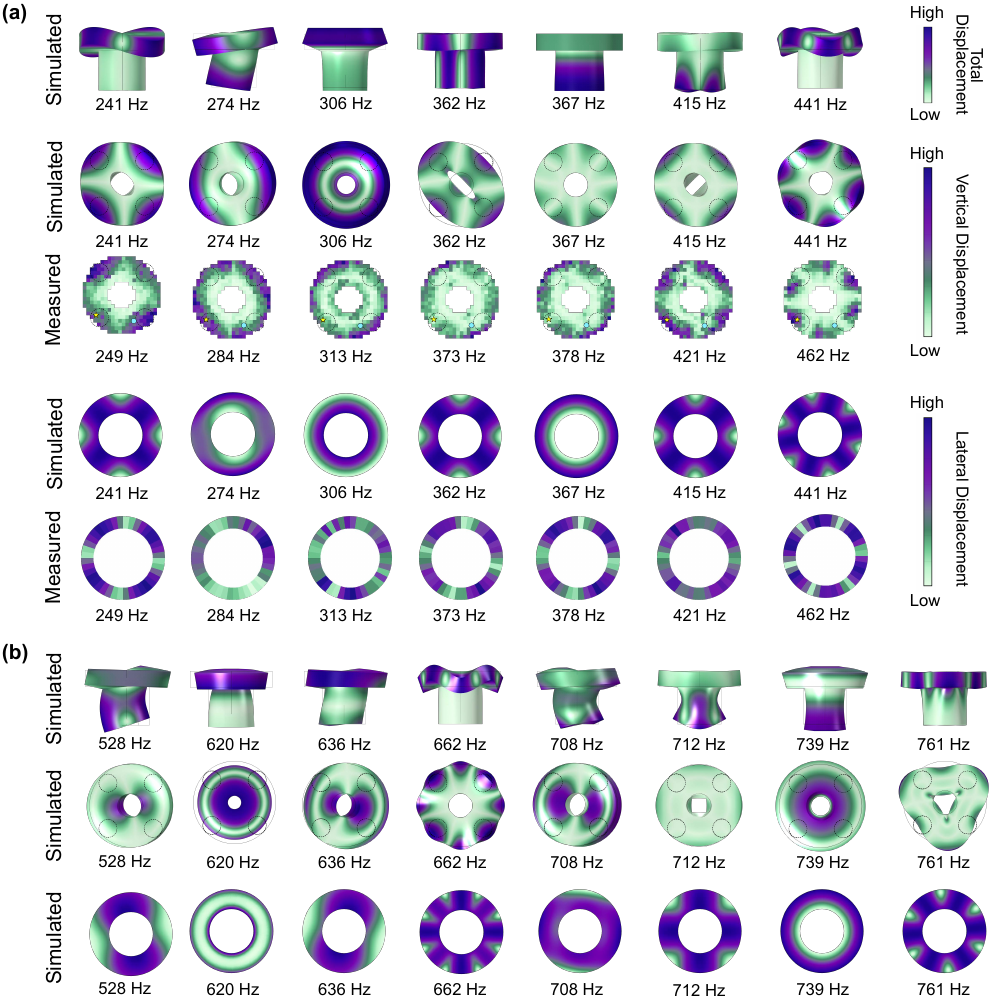}
    \caption{(a) The first seven simulated flexural resonance modes and corresponding measured displacements of the cylindrical block at these frequencies. In the first row, the purple-green colorscale shows total displacement, $\sqrt{(dx)^2+(dy)^2+(dz)^2}$. In the second pair of rows, the purple-green colorscale shows vertical displacement of the block's top face, $dz$. In the third pair of rows, the purple-green colorscale shows horizontal displacement of the block's tail, $\sqrt{(dx)^2+(dy)^2}$. Only the simulated modes at 241 Hz, 274 Hz, 306 Hz, and 441 Hz give rise to significant vertical displacement of the top surface of the cylindrical block. (b) The next eight simulated flexural resonance modes of the cylindrical block, showing total displacement, vertical displacement of the block's top face, and horizontal displacement of the block's tail, respectively, in the three rows. Although some of these modes are expected to give rise to significant vertical displacements, we did not measure any clearly identifiable modes above $\sim500$ Hz.}
    \label{fig:all-sim-modes}
\end{figure}

We simulated the flexural modes of our cylindrical inertia block using the commercial finite element analysis (FEA) package {\sc comsol multiphysics 5.4}. In practice, the allowed modes are partially constrained by the contacts with the pneumatic isolators. We simulated this constraint by creating circular domains representative of each isolator's contact area on the underside of the block. These circular surfaces were set to spring boundary conditions, with total effective spring constant $k=(2 \pi f_0)^2M$, where $f_0=1$ Hz is the resonance frequency of the isolator and $M=15,700$ kg is the mass of the block. The remainder of the block surface was set to free boundary conditions. We used the composite mass density 2300 kg/m$^3$ and Young's modulus $2.87\times10^{10}$ Pa, with Poisson's ratio 0.20. A complete set of simulated flexural modes up to 800 Hz is shown in Fig.\ \ref{fig:all-sim-modes}.

We measured the flexural modes of the cylindrical block using a Wilcoxon 731a accelerometer. A complete set of measured flexural modes up to 500 Hz is shown in Fig.\ \ref{fig:all-sim-modes}(a). Fig.\ \ref{fig5}(b) shows few measured spectral features above 500 Hz, which may be due to inefficiency of our mallet strike in exciting higher flexural modes of the cylindrical block, or due to limitations of the accelerometer itself. The nominal accelerometer sensitivity is 10 V/$g$, with accuracy $\pm 10\%$ from 0.1-300 Hz, and $\pm 3$ dB from 0.05-450 Hz. The accelerometer's own nominal resonance is 750 Hz, so measured amplitudes above this frequency are expected to be heavily suppressed.

\clearpage
\section{Measurements}

\begin{figure}[h!]
    \centering
    \includegraphics[width=\columnwidth]{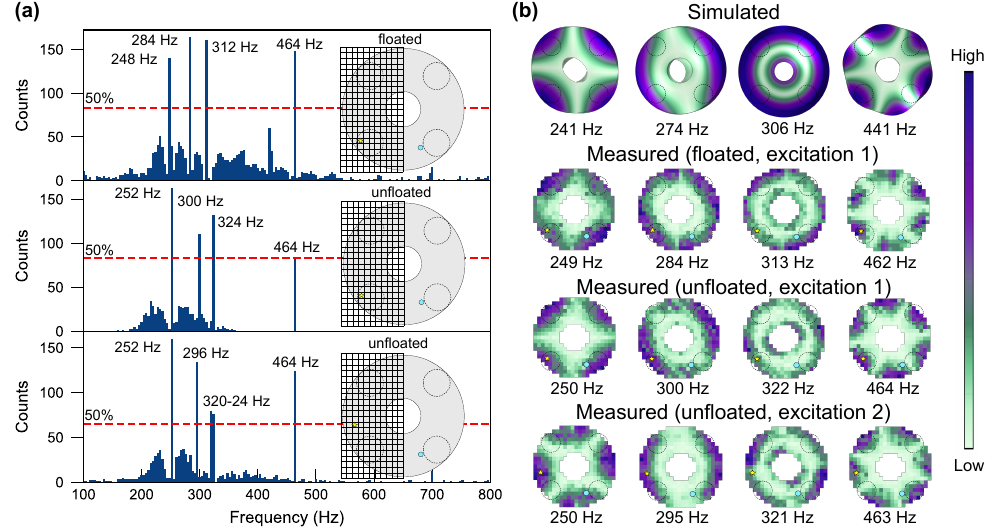}
    \caption{(a) Histograms of peaks in the vertical acceleration spectra of the block's top surface, when excited under different conditions. Top row is reproduced from main text Fig.\ \ref{fig5}(a), and shows the response when the block was floated and excited at the location marked by a yellow star (excitation 1). Second row shows the response when the block was unfloated but excited at the same location. Third row shows the response when the block was unfloated but excited at a different location (excitation 2). A slight upwards drift in the resonance frequencies of the four-fold and six-fold modes occurs in the unfloated datasets. (b) The first two rows show the simulated and measured top views of the first four flexural resonance modes that give rise to significant vertical displacement of the upper surface of the cylindrical block, reproduced from main text Fig.\ \ref{fig6}(b). The next row shows the measured mode shapes at the peak frequencies when the block is unfloated and excited at location 1. The last row shows the measured mode shapes at the peak frequencies when the block is unfloated and excited at location 2. The purple-green colorscale shows vertical displacement of the block's top face, $dz$.}
    \label{fig:second-data}
\end{figure}

During all measurements shown in main text Figs.\ 5 and 6, both excitation point and reference point were held constant, and the inertia block was floated, with the pneumatic isolators pressurized to 80 psi. Fig.\ \ref{fig:second-data} shows two analogous sets of measurements, with the block unfloated, and with a second excitation point.

\clearpage
\section{Q-factor}

The response function of a simple one degree of freedom (1DOF) driven mass-spring system is given by
\begin{equation}
    K(\omega) = \sqrt{\frac{\omega_{0}^4 + \omega^2\omega_0^2/Q^2}{(\omega_{0}^2 - \omega^2)^2 +  \omega^2\omega_0^2/Q^2}},
\label{eq:1DOF}
\end{equation}
where $\omega_0 = \sqrt{k/M}$, and $Q$ is the quality factor (inversely proportional to the damping of the spring).

\begin{figure}[h!]
    \centering
    \includegraphics{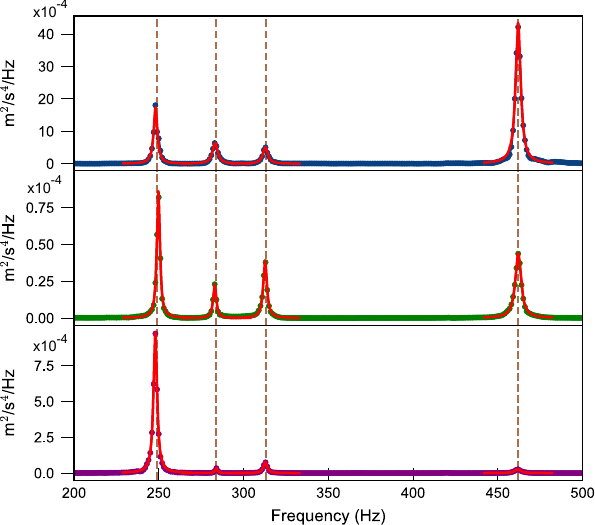}
    \caption{Lorentzian fits of peaks at 249 Hz, 284 Hz, 313 Hz, and 462 Hz shown for three spectra at various locations on the block surface, as defined in main text Fig.\ \ref{fig4}(a).}
    \label{fig:lorentzianfits}
\end{figure}

We assume that our mallet strike excites the cylinder with a $\delta$-function force, or equivalently acceleration, at an instant in time, so its frequency spectrum $a_{\mathrm{exc}}(\omega)$ is approximately flat. Therefore the measured acceleration response $a_{\mathrm{meas}}(\omega)$ is approximately proportional to the transfer function $K(\omega)$ in Eq.\ \ref{eq:1DOF}. The square of the response function, $|K(\omega)|^2$, reduces to a Lorentzian in the limit of $Q \gg 1$ and $|\omega-\omega_0| \ll \omega_0$,
\begin{equation}
    |K(\omega)|^2 \appropto \frac{1}{(\omega-\omega_0)^2 + \left( \displaystyle{\frac{\omega_0}{2Q}} \right)^2 },
    \label{eq:lorentzian}
\end{equation}
\noindent where $\omega_0/Q$ is the full width at half maximum (FWHM). Fig.\ \ref{fig:lorentzianfits} shows Lorentzian fits to three example power spectra, specifically the squared data from Fig.\ \ref{fig5}(a), on linear amplitude scale.

\begin{table}[h!]
\centering
\begin{tabular}
    {|r||c|c||c|c||c|c||c|c|}
 \hline
   \textbf{Peak} & \multicolumn{2}{c||}{\textbf{249 Hz}}
   & \multicolumn{2}{c||}{\textbf{284 Hz}}
   & \multicolumn{2}{c||}{\textbf{313 Hz}}
   & \multicolumn{2}{c|}{\textbf{462 Hz}} \\
 \hline
 & FWHM & $Q$ & FWHM & $Q$ & FWHM & $Q$ & FWHM & $Q$\\
 \hline
 \blue{\textbf{Blue}} & \blue{2.60 Hz} & \blue{96} & \blue{3.56 Hz} & \blue{80} & \blue{3.44 Hz} & \blue{91} & \blue{3.53 Hz} & \blue{131} \\
 \green{\textbf{Green}} & \green{2.14 Hz} & \green{117} & \green{2.13 Hz} & \green{133} & \green{2.54 Hz} & \green{123} & \green{4.02 Hz} & \green{115} \\
 \purple{\textbf{Purple}} & \purple{2.47 Hz} & \purple{100} & \purple{1.66 Hz} & \purple{171} & \purple{2.59 Hz} & \purple{121} & \purple{4.30 Hz} & \purple{107} \\
 \hline
 \textbf{Avg (165 locs)} & 2.66 Hz & 103 & 3.24 Hz & 122 & 3.14 Hz & 106 & 4.10 Hz & 119 \\
 \hline
\end{tabular}
\caption{Lorentzian fits to $|a(f)|^2$ to determine $Q_{\mathrm{concrete}}$.}
\label{table:Qfactor}
\end{table}

\clearpage
\section{Transfer Functions}

A low-vibration laboratory is designed to reduce the transmission of typical building vibrations to a sensitive instrument. In this case, we consider the instrument to be a scanning tunneling microscope (STM). The design effectiveness can be quantified by the transfer function: the frequency spectrum of the tip-sample displacement, divided by the frequency spectrum of the building noise. The overall transfer function can be decomposed as the product of several components.

First, the heavy block on pneumatic isolators acts as a filter to attenuate high-frequency noise above a cutoff that depends on the mass of the block ($M$) and the stiffness of the pneumatic isolators ($k$). Meanwhile, damping absorbs energy so that vibrations decay more quickly in time. The damping of the pneumatic isolators depends on the constituent material properties and their friction, and is typically engineered for quality factor $Q_{\mathrm{pneumatic}} \approx 3-10$ to achieve optimal balance of fast rolloff of the filter vs.\ low amplification of the rigid body modes of the floating block at the pneumatic isolator resonance.\cite{Chen2007} The transfer function for a simple one degree of freedom (1DOF) mass-spring system is given by Eq.\ \ref{eq:1DOF}. After a peak in which the incoming vibrations are amplified by a factor $Q$, the transfer function falls off as $1/f^2$ until $f \sim Qf_0$, and then continues to fall off more slowly as $1/fQ$.

\begin{figure}[h!]
    \centering
    \includegraphics{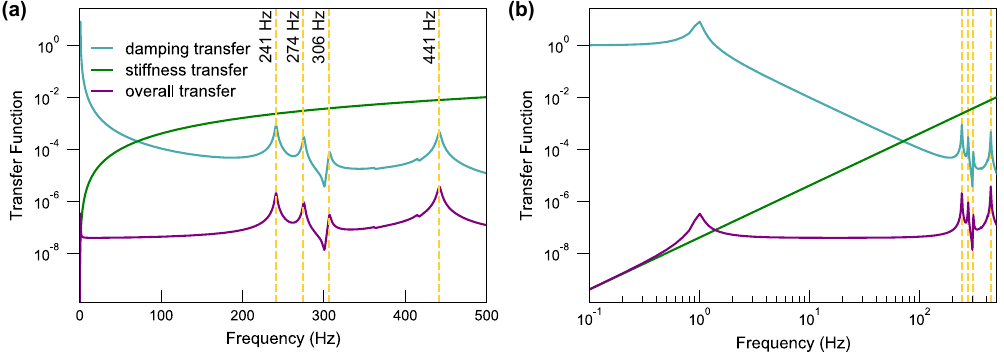}
    \caption{Linear and logarithmic plots of the damping, stiffness, and overall transfer functions for an STM on a floating concrete cylinder. The damping transfer function was simulated with FEA using {\sc comsol}, with pneumatic isolator properties $f_0 = 1$ Hz and $Q_{\mathrm{pneumatic}}=10$, and concrete properties mass density 2300 kg/m$^3$, Young's modulus $2.87\times10^{10}$ Pa, Poisson's ratio 0.20, and damping $\eta_{\mathrm{concrete}} = 0.008$. The stiffness transfer function derives from a simple 1DOF mass-spring model, is calculated for STM properties $Q_{\mathrm{STM}} = 100$ and $f_{\mathrm{STM}} = 5$ kHz. The overall transfer function is a product of the damping and stiffness transfer functions.}
    \label{fig:pohl}
\end{figure}

Second, the components of the isolation system (e.g.\ the floating concrete block, or the frame on which the STM sits) may have flexural modes of their own, which contribute peaks to the transfer function at their resonance frequencies. We used {\sc comsol} to simulate the full ``damping transfer'' function, including the rigid body mode of the cylindrical block in Eq.\ \ref{eq:1DOF} (with $f_0 = 1$ Hz and $Q_{\mathrm{pneumatic}} = 10$) and its flexural resonances (with $\eta_{\mathrm{concrete}} = 0.008$). The result is shown as the blue curve in Fig.\ \ref{fig:pohl}.

Third, we are interested in the relative displacement of the tip and sample in the STM. The STM itself can be modeled as a driven mass-spring system, where the tip and sample sit on different plates that are connected by a stiff spring (the body of the STM), which is engineered to be as rigid as possible. Following Ref.\ \onlinecite{PohlIEEE1986}, we model the ``stiffness transfer'' of the STM with $f_{\mathrm{STM}} = 5$ kHz and $Q_{\mathrm{STM}} = 100$, and show the result as the green curve in Fig.\ \ref{fig:pohl}.

Finally, the overall transfer function is the product of the damping and stiffness transfer functions, shown as the purple curve in Fig.\ \ref{fig:pohl}. We see that the flexural modes of the concrete cylinder feature prominently, raising the overall transfer function by more than an order of magnitude over its background, to around $10^{-6}$. It is therefore important to ensure that these peaks lie at frequencies above the expected building vibration peaks shown in main text Fig.\ \ref{fig1}.

\end{document}